# Resonant Bloch-wave beatings

Yaroslav V. Kartashov,[1,2,*] Victor A. Vysloukh,[3] and Lluis Torner[1]

[1]*ICFO-Institut de Ciencies Fotoniques, and Universitat Politecnica de Catalunya, Mediterranean Technology Park, 08860 Castelldefels (Barcelona), Spain*
[2]*Institute of Spectroscopy, Russian Academy of Sciences, Troitsk, Moscow Region, 142190, Russia*
[3]*Departamento de Fisica y Matematicas, Universidad de las Americas - Puebla, Santa Catarina Martir, 72820, Puebla, Mexico*

We introduce Bloch-wave beatings in arrays of multimode periodically bent waveguides with a transverse refractive index gradient. The new phenomenon manifests itself in the periodic drastic increase of the amplitude of the Bloch oscillations that accompanies resonant conversion of modes guided by the individual waveguides. The Bloch-wave beatings are found to be most pronounced when the length of the resonant mode conversion substantially exceeds the longitudinal period of the Bloch oscillations. The beating frequency decreases when the amplitude of waveguide bending decreases, while the beating amplitude is restricted by the amplitude of the Bloch oscillations that emerge from the second allowed band of the Floquet-Bloch lattice spectrum.

Today the phenomena of Bloch oscillations and Zener tunneling, that were predicted several decades ago in solid-state physics [1,2], are widely known in diverse areas of science. They appear as electronic oscillations in semiconductor superlattices [3], in matter waves in optically-trapped ensembles of ultracold atoms [4], and even in acoustic fields in layered structures [5]. Photonic lattices (or waveguide arrays) are a powerful physical setting to explore optical analogues of a wide variety of quantum-inspired effects, including Bloch oscillations. In the presence of a transverse gradient of the refractive index, the eigenmodes of otherwise periodic optical waveguide arrays transform into localized Wannier-Stark modes and their propagation constants become equidistant, forming so-called Wannier-Stark ladders [6]. In such a system periodic revivals of input light beam occur after a certain propagation distance.

Optical manifestations of Bloch oscillations and Zener tunneling have been observed and extensively studied in both one- and two-dimensional waveguide arrays [7-12] (see also [13-15] for recent reviews on light evolution in waveguide arrays, and references therein). Recently, so-called fractional Bloch oscillations were revealed [16] and transition from Bloch oscillations to Anderson localization in disordered waveguide arrays was experimentally addressed in [17]. A weak nonlinearity destroys the periodicity of Bloch oscillations [9,11], and formation of unusual solitons, emerging from the localized Wannier-Stark linear modes, may occur for a sufficiently strong nonlinearity [18].

All such previous investigations of Bloch oscillations have addressed only waveguiding structures consisting of single-mode waveguides, which thus cannot feature internal mode-conversion inside the individual guides. However, if the array consists of multimode waveguides, the rate of light energy transport across the structures and the amplitude of the collective Bloch oscillations may strongly depend on the particular superposition of the excited individual modes, due to difference in the overlap integrals of different modes with neighboring waveguides [19]. In that scenario, resonant mode conversion utilizing longitudinal waveguide modulation [15], affords rich possibilities for the dynamical control of the weights of guided modes and, therefore, the global diffraction rate [20,21]. Different approaches are available to realize resonant mode conversion, including periodic mechanical stress and photo-induced gratings [22-24].

In this Letter we aim to study how the internal mode conversion in multimode waveguides in longitudinally-modulated arrays affects the dynamics of Bloch oscillations. We find that the longitudinal modulation dynamically couples the excitations emerging from the first and second allowed bands and leads to Bloch-wave beatings, which manifest themselves in periodic drastic increase of the amplitude of the Bloch oscillations.

The propagation of a light beam along the $\xi$-axis of a medium with transversally and longitudinally inhomogeneous refractive index landscape and Kerr-type nonlinearity is described by the nonlinear Schrödinger equation for the dimensionless light field amplitude $q$:

$$i\frac{\partial q}{\partial \xi} = -\frac{1}{2}\frac{\partial^2 q}{\partial \eta^2} - \sigma q|q|^2 - pR(\eta,\xi)q - \alpha\eta q, \qquad (1)$$

where $\eta$ is the transverse coordinate normalized to the characteristic transverse scale $x_0$; $\xi$ is the propagation distance normalized to the diffraction length $k_0 x_0^2$; $p = \delta n k_0^2 x_0^2 / n$ is the waveguiding parameter that is proportional to the refractive index modulation depth $\delta n \ll n$; $k_0$ is the wavenumber, the parameter $\sigma > 0 / \sigma < 0$ corresponds to a focusing/defocusing nonlinearity, while $\alpha$ stands for the transverse refractive index gradient. To describe the refractive index profile we use a superposition of super-Gaussian functions $R(\eta,\xi) = \sum_{j=-\infty}^{\infty} \exp\{-[\eta - \mu\sin(\Omega\xi) - jd]^4 / a^4\}$, where the parameter $\mu$ describes the amplitude of the in-phase bending of all waveguides in the array; $\Omega$ is the bending frequency; $d$ is the period of the array, and $a$ is the waveguide width. In this paper we set the bending ampli-tude $\mu \ll 1$. Longitudinal modulation periods $2\pi/\Omega$ are assumed to be much larger than the wavelength of light, therefore one may treat the waveguide array as a long-period grating [22-24]. We set $d=2$, $a=0.5$ and consider first the case of linear media with $\sigma=0$. The depth $p=14$ of the individual waveguide in selected in such a way that only two guided modes are supported, in the form of fields which are written as $q_k(\eta,\xi)=w_k(\eta)\exp(ib_k\xi)$, where $k=1,2$, with propagation constants (eigenvalues) $b_1=10.525$, $b_2=2.924$ at $\alpha=0$.

The spatial frequency of the resonant mode conversion in a single waveguide can be elucidated using standard mode coupling technique, which is based on the modal expansion $q(\eta,\xi)=c_1(\xi)w_1(\eta)e^{ib_1\xi}+c_2(\xi)w_2(\eta)e^{ib_2\xi}$ of the total field. Substitution of this expansion into Eq.(1) yields a system of two ordinary differential equations describing the evolution of the complex coefficients $c_k(\xi)$ that is reminiscent of the evolution equations for populations in a two-level driven quantum system:

$$dc_1/d\xi = -\mu A_{12} e^{i(\Omega-\Omega_{12})\xi} c_2/2, \quad (2)$$
$$dc_2/d\xi = \mu A_{21} e^{-i(\Omega-\Omega_{12})\xi} c_1/2,$$

where $\Omega_{12}=|b_1-b_2|$, and the exchange integrals $A_{ik}=p\int_{-\infty}^{\infty}w_i(\eta)R_0'(\eta)w_k(\eta)d\eta$ involve the derivative of the waveguide profile $R_0'=\partial R/\partial\eta|_{\xi=0}$. The presence of a longitudinal bending stimulates the coupling between guided modes of different parity $(w_1 \leftrightarrow w_2)$ because corresponding exchange integrals contain the anti-symmetric kernel terms $R_0'(\eta)=-R_0'(-\eta)$. At resonance $(\Omega \approx \Omega_{12})$, the frequency of the periodic energy exchange between guided modes is given by $\Omega_c=\mu A_{12}/2$ and it can be controlled by varying the bending amplitude $\mu$.

If one has an infinite waveguide array composed from unmodulated two-mode waveguides and the parameter $\alpha=0$, one observes the formation of a band-gap structure, where the first and second allowed bands in the Floquet-Bloch spectrum have the borders $b_1^{\rm upp}=10.546$, $b_1^{\rm low}=10.504$ and $b_2^{\rm upp}=3.245$, $b_2^{\rm low}=2.491$, respectively. It should be properly appreciated that these bands are centered on the propagation constants of the first and second modes supported by the individual waveguides. The eigenmodes of such an array are Bloch waves, termed $w_B(\eta)$. In the presence of longitudinal bending resonant coupling can be achieved between Bloch waves [25]. The mathematical description of this effect is similar to the coupled-mode approach used above for a single waveguide. The resonant frequency at which coupling between bands occurs remains close to the value $\Omega_{12}$.

However, in the presence of a transverse refractive index gradient $(\alpha \neq 0)$ the spectrum changes qualitatively, with important physical consequences. Figure 1(a) shows the transformation of a representative eigenvalue spectrum of a *finite* array containing 95 waveguides when the gradient $\alpha$ increases. Only positive eigenvalues $b_k$ are shown. At small gradients, such as $\alpha=0.02$, the eigenvalues are still grouped into two well-separated bands centered around the eigenvalues of the isolated waveguide (dashed lines). In this case the spectrum resembles a band-gap structure of a fully periodic array. When $\alpha$ increases, the bands broaden and finally merge (see curve 3, calculated for $\alpha=0.04$). Then the spectrum is composed out of equidistantly-spaced eigenvalues with $b_k-b_{k+1}=\alpha d$. The deviation from linear dependence is due to the fact that we study a *finite* array where some modes can be affected by the presence of the array boundaries. The eigenmodes of such an array are localized Wannier-Stark modes. Examples of modes centered at $\eta=0$ are shown in Fig. 1(b). Because the array is composed of two-mode waveguides, there are *two* types of modes whose internal structure resembles the fundamental and second modes of the single waveguides. The mode whose eigenvalue is close to $b_2$ is significantly more extended than the mode with $b_k \approx b_1$.

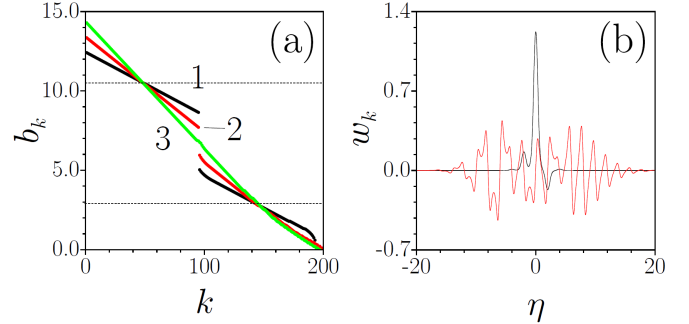

Fig. 1. (a) Eigenvalue spectrum of a finite array with 95 waveguides at $\alpha=0.02$ (black circles, 1), $\alpha=0.03$ (red circles, 2), and $\alpha=0.04$ (green circles, 3). Horizontal dashed lines correspond to propagation constants of the first and second guided modes of the individual waveguides at $\alpha=0$. (b) Profiles of eigenmodes centered at $\eta=0$ with eigenvalues close to the upper (black curve) and lower (red curve) dashed lines in (a).

Since the Wannier-Stark ladder involves a number of localized modes with a small propagation constant offset, the standard coupled-mode approach is ineffective for describing the longitudinally modulated arrays incorporating a gradient. We thus resort to numerical integration of Eq. (1). We used input conditions in the form of the Bloch waves associated to arrays with $\alpha=0$, embedded in broad Gaussian envelopes, which can be written as $q(\eta,0)=A\exp(-\eta^2/W^2)w_B(\eta)$, where $W \gg d$. The input amplitude was adjusted so that $U=\int_{-\infty}^{\infty}|q(\eta,0)|^2 d\eta=1$. Without bending such inputs give rise to standard periodic Bloch oscillations, whose longitudinal period is given by $\xi_B=2\pi/\alpha d=2\pi/(b_k-b_{k+1})$. However, the amplitude of oscillations strongly depends on the internal structure of the beam. It can be estimated as $\eta_B \approx (b_k^{\rm upp}-b_k^{\rm low})/\alpha$, where $k$ denotes the number of the band in the Floquet-Bloch spectrum associated to $w_B(\eta)$.

Figures 2(a) and 2(b) illustrate oscillations of the integral center of the beams $\eta_c(\xi)=U^{-1}\int_{-\infty}^{\infty}\eta|q(\eta,\xi)|^2 d\eta$ originating from different bands and for different refractive index gradients $\alpha$ in the absence of longitudinal modulation. We used an envelope with $W=7.6$. The profiles of Bloch modes $w_B(\eta)$ were selected from the top of the first and second bands ($b=b_1^{\rm upp}$ and $b=b_2^{\rm upp}$). Since the second band ($b_2^{\rm upp}-b_2^{\rm low}=0.754$) is remarkably wider than the first band ($b_1^{\rm upp}-b_1^{\rm low}=0.042$), the amplitude of Bloch oscillations for the second-band excitations exceeds that for the first-band excitations nearly by a factor of 18, as readily visible in curves 1 and 2 of Fig. 2(a). The corresponding evolution dynamics is illustrated in Figs. 3(a) and 3(b), where the first-band excitation experiences only very small displacement, while its second-band counterpart strongly oscillates. The amplitude of oscillations rapidly grows when the refractive index gradient becomes smaller [Fig. 2(b)].

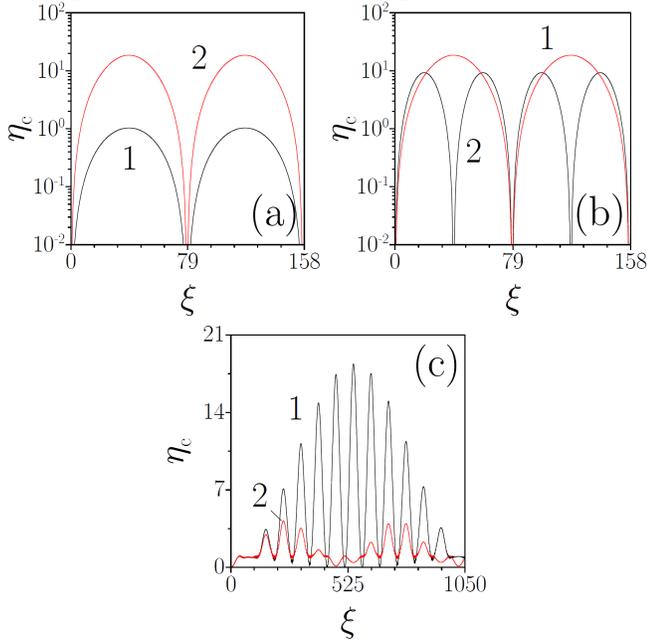

Fig. 2. (a) Integral beam center $\eta_c$ versus propagation distance $\xi$ for $\alpha = 0.04$ and input beams originating from the first (curve 1) and second (curve 2) bands at $\mu = 0$. (b) $\eta_c$ versus $\xi$ for the input beam originating from the second band at $\mu = 0$ and $\alpha = 0.04$ (curve 1), $\alpha = 0.08$ (curve 2). (c) $\eta_c$ versus $\xi$ in the presence of longitudinal modulation at $\alpha = 0.04$, $\mu = 0.001$ and $\nu = 0$ (curve 1), $\nu = 0.15$ (curve 2). Note the logarithmic scales in (a) and (b).

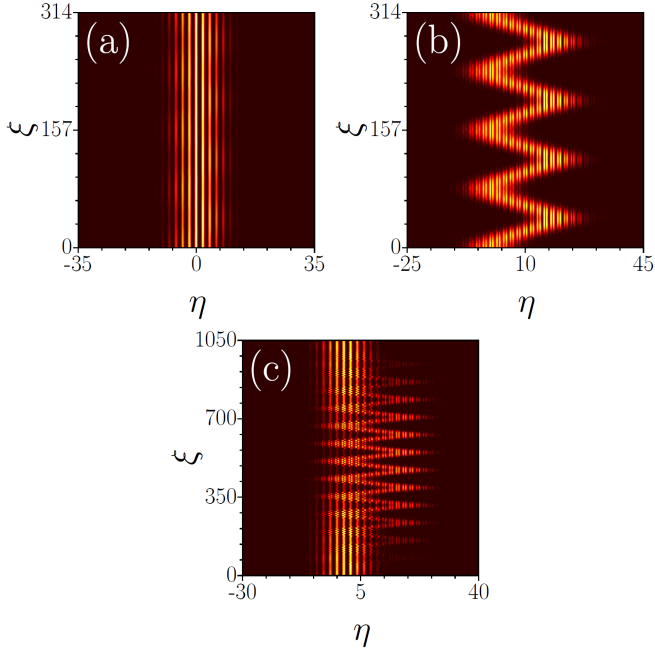

Fig. 3. Evolution dynamics in an array without longitudinal modulation in individual channels when the input wavepacket originates from the first (a) and second (b) bands at $\alpha = 0.04$. (c) Evolution dynamics in a modulated array with $\Omega = 7.335$, $\mu = 0.001$, $\alpha = 0.04$ for the input wavepacket originating from the first band.

A longitudinal bending of the waveguides qualitatively changes the character of Bloch oscillations, provided that the frequency of the longitudinal refractive index modulation is sufficiently close to its resonant value. We introduce the normalized frequency detuning $\nu = (\Omega - \Omega_r)/\Omega_r$, where the resonant beating frequency is given by $\Omega_r = b_1^{\text{upp}} - b_2^{\text{upp}}$, which is a quantity close to $\Omega_{12}$ for resonant coupling in isolated waveguides. The longitudinal bending gradually changes the internal structure of the excitation – after a certain propagation distance the field distribution inside each waveguide starts resembling the second guided mode rather than the first one. This leads to an increase of the tunneling rate and to a larger light deflection in the direction of the refractive index gradient. Note that this is in full agreement with a larger amplitude of the Bloch oscillations for the second-band excitations in the unmodulated arrays.

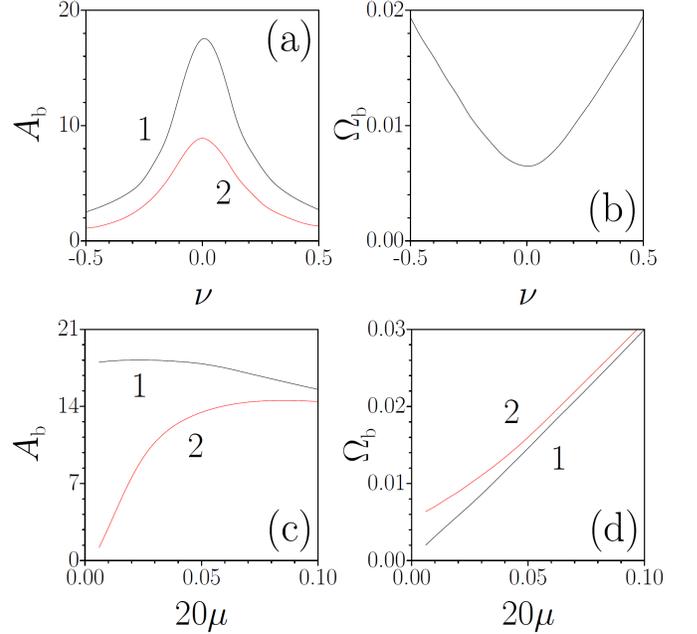

Fig. 4. (a) Beating amplitude versus detuning for $\alpha = 0.04$ (curve 1) and $\alpha = 0.08$ (curve 2) at $\mu = 0.001$. (b) Beating frequency versus detuning at $\alpha = 0.08$, $\mu = 0.001$. (c) Beating amplitude (c) and frequency (d) as functions of modulation depth for $\nu = 0$ (curves 1) and $\nu = -0.1$ (curves 2) at $\alpha = 0.04$.

However, since resonant coupling is a periodic process, upon subsequent propagation, the first-band structure of the excitation is restored and the amplitude of Bloch oscillations decreases again to its initial value. As a result one observes periodic *Bloch-wave beatings*, as clearly visible in the $\eta_c(\xi)$ dependence in Fig. 2(c) and in the propagation dynamics from Fig. 3(c). It should be stressed that this picture is observed as long as the distance of resonant coupling substantially exceeds the period $\xi_B$ of Bloch oscillations, since it is this coupling distance that determines the periodicity of the beatings. At exact resonance ($\nu = 0$) the amplitude of beatings $A_b(\xi)$ is a periodic function of the propagation distance. Its maximal value is limited by the amplitude of the Bloch oscillations for the second-band excitations, while the minimal amplitude is dictated by $\eta_B$ for the first-band excitations [see Fig. 2(c), curve 1]. Even small frequency detuning ($\nu = 0.15\%$) reduces the period of Bloch beatings by a factor of two and drastically decreases their amplitude, as shown in Fig. 2(c), curve 2. We observed that narrower wavepackets with $W=3.5$ exhibit similar evolution, but with slightly reduced amplitude of beatings.

Figure 4 illustrates the resonant nature of the Bloch-wave beatings phenomenon. The amplitude $A_b$ of the beatings is found to strongly depend on the relative frequency detuning $\nu$. The resonances are narrow [Fig. 4(a)] and already a small 1% detuning destroys the effect. Increasing the refractive index gradient $\alpha$ reduces the beating amplitude, so that the effect is most pronounced at small $\alpha$. The beating frequency acquires its minimal value exactly at the resonant modulation frequency [Fig. 4(b)] and it only weakly depends on the gradient $\alpha$. The dependence of the beating amplitude on the amplitude of the waveguide bending is shown in Fig. 4(c), and it is unusual: while at resonance (i.e., $\nu=0$, curve 1) the beating amplitude only slightly decreases with $\mu$ (an effect mainly due to increase of radiative losses), in the case of non-resonant modulation (i.e., $\nu=-0.1$, curve 2) the beating amplitude notably grows with $\mu$ and eventually saturates. On the other hand, the beating frequency is a monotonically growing function of the bending amplitude for any detuning [Fig. 4(d)].

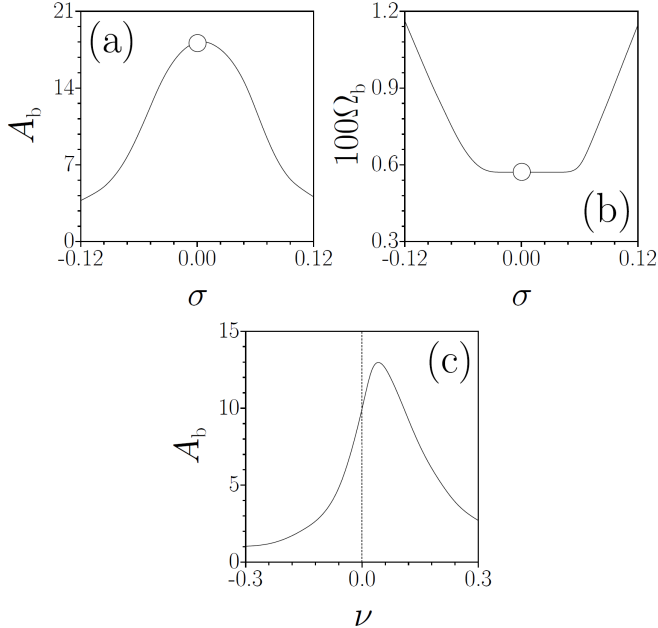

Fig. 5. Beating amplitude (a) and frequency (b) vs nonlinearity strength at $\alpha=0.04$, $\mu=0.001$, $\nu=0$. Circles correspond to evolution dynamics shown in Fig. 3(c). (c) Beating amplitude versus detuning at $\sigma=0.07$, $\alpha=0.04$, and $\mu=0.001$.

We found that a material nonlinearity strongly affects Bloch-wave beatings. Upon increase of the absolute value of the nonlinearity parameter $\sigma$ the amplitude of beatings diminishes [Fig. 5(a)], while the frequency of beatings grows [Fig. 5(b)]. Increasing $\sigma$ leads to the appearance of an additional nonlinear phase mismatch between eigenmodes (the variation of the propagation constant of the first mode of the isolated waveguide due to the nonlinearity is larger than that of the second mode), which drives the system away from resonance and thus delays the process of transformation of the internal structure of the wavepacket. Note that due to the intensity difference in the excited guides, resonant mode conversion is firstly suppressed in the central channels, leading to the spatially nonuniform freezing of the Bloch-wave beatings. Interestingly, the dependence of the beating frequency on $\sigma$ features an extended plateau. Nonlinearity leads to notable asymmetries in the resonant curves, as depicted in Fig. 5(c), which was obtained for the focusing Kerr nonlinearity. We found that while in a focusing medium the resonance frequency increases, in a defocusing medium it decreases.

In conclusion, we introduced the phenomenon of Bloch-wave beating, which takes place under suitable conditions in arrays of harmonically-bent bimodal waveguides with a built-in transverse refractive index gradient.